# Disentangling the schema turn:
# Restoring the information base to conceptual modelling

Chris Partridge[1, 2] [0000-0003-2631-1627], Andrew Mitchell[1, 2] [0000-0001-9131-722X], Sergio de Cesare[2] [0000-0002-2559-0567] and Oscar Xiberta Soto[1] [0000-0002-0324-0726]

[1] BORO Solutions Ltd., London, UK
[2] University of Westminster, London, UK
`partridgec@boro.solutions`

**Abstract.** If one looks at contemporary mainstream development practices for conceptual modelling in computer science, these so clearly focus on a conceptual schema completely separated from its information base that the conceptual schema is often just called the conceptual model. These schema-centric practices are crystallized in almost every database textbook. We call this strong, almost universal, bias towards conceptual schemas the *schema turn*.

The focus of this paper is on disentangling this turn within (computer science) conceptual modeling. It aims to shed some light on how it emerged and so show that it is not fundamental. To show that modern technology enables the adoption of an inclusive schema-and-base conceptual modelling approach, which in turn enables more automated, and empirically motivated practices. And to show, more generally, the space of possible conceptual modelling practices is wider than currently assumed. It also uses the example of bCLEARer to show that the implementations in this wider space will probably need to rely on new pipeline-based conceptual modelling techniques. So, it is possible that the schema turn's complete exclusion of the information base could be merely a temporary evolutionary detour.

## 1 Introduction

Conceptual model (as a term, at least) first emerged in the 1970s and 80s in the context of computing and its original sense was a representation of the 'real' world. Soon afterwards this notion of a model was divided into the conceptual schema and the information base [1] – these divided senses are preserved in some more recent texts (e.g. [2]). (Here, we will use the historically accurate term *information base*, sometimes shortened to *base*, rather than the looser terms, *instance* or *data*.) If one looks at contemporary mainstream development practices for conceptual modelling in computing, then there is a strong, almost universal, bias towards conceptual schemas. The focus is on a conceptual schema completely separated from its information base – to the extent that the conceptual schema is often just called the conceptual model. In a typical development life cycle, the information base appears at the end of the cycle, as physical data, as a physicalized version of the information base, and is added to the physical system during implementation. This is long after the main work of real-world conceptual modelling done in the early stages. This schema-centric practice of separating and shifting the information base is crystallized in almost every database textbook (e.g. [3]).

The schema-centric practice within conceptual modelling is particularly apparent to us as, in our ontological practice, we bind the schema and base together, working with an inclusive schema-and-base conceptual model [4]. This naturally raises questions for us about why this practice emerged and how fundamental it is to conceptual modelling.

### 1.1    What is not in scope

The use of conceptual modelling (or maybe more accurately the use of the term) has spread across a wide range of disciplines – see, for example, such disparate uses as hydrology [5] and psychiatry [6]. This wider use of the term is outside the scope of the paper which is restricted to computer science and information systems development. Thalheim [2] calls this context 'computer science conceptual modeling'. Hence, for the rest of the paper, when we refer to conceptual modeling we mean 'computer science conceptual modeling'.

There is also a wider related notion of conceptual scheme. This can be traced back to at least Kant's categories of understanding – in [7] he used the German word 'schema'. It continues in Carnap [8] and Quine [9] – even Kuhn's [10] paradigms. Davidson's critique of the conceptual scheme [11] was written in the early 1970s, around the same time as the computer science conceptual schema work emerged. What is intriguing is that he suggests (in his context) that "the dualism of scheme and content, of organizing system and something waiting to be organized, cannot be made intelligible and defensible." Though clearly related, this wider notion is unfortunately also outside the scope of this paper.

### 1.2    What is in scope

The focus of the paper is on what we the *schema turn* within (computer science) conceptual modeling. It aims firstly to shed some light on how this turn emerged. It then aims to show that the turn is not fundamental, explaining how modern technology enables the adoption of an inclusive schema-and-base conceptual modelling approach. This shows more generally that the space of possible conceptual modelling practices is wider than currently assumed. It also raises the possibility that the turn's complete exclusion of the information base is merely a temporary evolutionary detour.

## 2 Paper structure

When we looked in more detail at how and why the turn happened, we found a complicated situation. The evolution of the conceptual model, the division into schema and base and separation of the information base as well as our inclusive structures needed a more considered analysis than we expected. We found we needed a couple of intertwined frameworks to help us make overall sense of it.

The first framework was a 'language' with the right kind of expressivity to enable us to talk about the separating of the conceptual schema – and information base's subsequent unification. It was clear that architectural modularity was central – so we constructed a way to talk about this – based upon what we call modularity architectural styles. The second framework was adopting an existing information evolutionary context, one suitable for framing the emergence of conceptual modelling and its subsequent modularity architectural style adaptations.

Hence, in the first part of the paper, we start by introducing the two frameworks. Then we use these to look backwards at the emergence of conceptual modelling, building up a picture of the overall evolutionary landscape based upon modularity architectural styles. The evolutionary perspective helps us to appreciate the historic environmental pressures that led to the current situation, while noting these have radically changed. We also use what we have learnt to fill out the modularity architectural style framework.

Building on this work, in the second part of the paper, we look forwards. The modularity framework, and the landscape it reveals, provide us with a useful tool to explore architectural possibilities. To illustrate this, we use it to look at the possible modularity architectural styles for a common type of project – package application migration. And we show that the information base has a central role and highlight the benefits of a modularity architectural style which aggregates the information base and schema.

We have been working now for decades with this kind of modularity architecture, implementing it in a data pipeline based upon a framework now called bCLEARer. The paper describes the bCLEARer pipeline as an example of an implementation of a schema-and-base model. It introduces the pipeline approach by contrasting this with the way the schema turn shapes current conceptual modelling tools and associated practices. The schema turn's schema bias led to CASE tools (which evolved into modern UML tools) that use manual graphically based editing focused on the schema. This schema bias persisted through the emergence of Model-Driven Engineering (MDE) and textual Domain Specific Language (DSL) tools. The paper discussed how the return to a schema-and-base approach requires the sort of pipeline approaches developed for data engineering. Ones that enable more automated and empirically motivated practices. We briefly describe the practical lessons we have learnt. We then outline possible future work and finally summarize the paper.

## 3     Developing a framework for modularity

Modularity is a topic in many disciplines [12], [13], [14]. It is seen as an integral feature of complex systems [15]. It is a common theme in evolution where it is described as nature's way of dealing with complexity [16], [17]. So, it should not be surprising that separating the information base and conceptual schema into two or integrating them into one within a conceptual model turns out to be ways of choosing a modularity architecture.

As mentioned earlier, we found that for our analysis we needed a language, a framework, a way to talk about and disentangle these choices. We needed a way to characterize the broad structure of modularity in the evolutionary space of computer applications and more specifically, conceptual modelling. We built this by mixing two components.

The first component is Shaw and Garlan's [18] notion of an architectural style in software architecture. This gives us a sufficiently general way to talk about architectural modularity, one within which we can characterize our design space in terms of modularity architectural styles – where an architectural style characterizes a family of computer applications in terms of a pattern of modular structural organization. Each style can be seen as taking a specific perspective over the space of possible ways of organizing. Hence, an application's architecture can be categorized using multiple styles each using a different perspective. These styles represent the most coarse-grained level of application organization.

As Shaw and Garlan [18] point out, these styles are often implicit and so implemented intuitively. One of the common challenges which we also face is identifying the styles, making them explicit and so more amenable to engineering. For this we developed our second component, a hylomorphic framework. This enables us to separate the analysis of the modular form from the content. It also lets us use mixing as a measure of integration – this is more suited to architectural work than the usual software engineering measures that focus on interfaces [19]. To give some context, in the next section we describe the original Aristotelian hylomorphic framework. We then, in the subsequent section, describe our adaptation of it.

### 3.1    The original hylomorphic framework

Hylomorphism was originally proposed by Aristotle [20] as a metaphysical framework which explained the integration of matter (ὕλη, *hylē*) and form (μορφή, *morphē*) into substance (οὐσία ousia). Within the general division of form and matter, Aristotle used a process of division (διαίρεσις, *diairesis*) to develop hierarchies.

Being truly mixed is an important feature of all form and matter mixtures – and some substance mixtures. For Aristotle, a true mixture (*mixis*), unlike a mere mechanical aggregation, is a new homogeneous substance with a single, unified form. This kind of distinction between true mixing and mere aggregation may already be familiar to some from UML's roughly similar distinction between composition and aggregation.

### 3.2 A hylomorphic language for modularity

We have found the broad hylomorphic framework a useful, pragmatic way to organize the modularity architectural styles and talk about the modularity architecture of the application. We have different goals from Aristotle, who was aiming for a metaphysical theory about the (whole) world. We are only pragmatically using the framework to organize a small aspect (modularity) of a small portion of the modern world (computer applications). Hence, we do not need to adopt hylomorphism in any deep metaphysical sense.

We adopt broadly similar structures. We similarly start with a division into formal and material styles, where the (pure) formal styles capture the 'general' modularity structure and the material styles capture the types of distinction that motivate the modularity. (We make the common term shift from 'form' to 'formal' and 'matter' to 'material', as this makes more sense here.) We recognize that the computer applications involve representation, and so indirectly involve the represented. For the formal styles' modules, we consider only the form of the representation. But for the material styles' distinctions, we are interested in both the representation and the represented – or, at least, what we take as being represented. Conceptual modelling can be seen as the representation inheriting distinctions from the represented.

And similarly, both the formal and material styles are further divided, with different divisions for different purposes, and organized into hierarchies. In hylomorphic fashion, when a general formal module style is mixed with a material distinction style, it is always truly mixed – a *mixis* – and always produces a concrete modularity style (by analogy with Aristotle's substance). The concrete modularity styles are in a hierarchy mostly inherited from their formal and material component hierarchies. The result is a recognizably hylomorphic structure.

Given we have a different scope from Aristotle, we need to be clear what mixing is on our context. We adopt a simple basis for categorizing mixing. This simpler basis is the extensional separation of the representation (the data objects), in other words that they can be physically separated in space and/or time. Based on this pragmatic notion of separation, within the paper we identify three broad categories of mixing for analyzing modularity architectural styles – giving us a simple taxonomy:

- separated: a complete separation with no mixing,
- aggregated: analogous to an Aristotelian aggregate,
- integrated: analogous to an Aristotelian *mixis* (true mixture).

A good example of a formal style is Dijkstra's separation of concerns [21]. This is formal in the required sense of not being about any specific material distinction. An example of a material distinction is the topic of the paper, the now familiar division of the conceptual model into the conceptual schema and information base (which we will examine later). This is just a distinction with no commitment to any formal modularity architecture, though there is the potential for various architectures. If we hylomorphically mix the two then we get the separation of the conceptual schema from the information base – a concrete modularity architectural style. This is likely to be part of a broader style that handles the eventual unifying of the conceptual schema and information base.

This example illustrates the dual role of mixing. The hylomorphic theme of separating and mixing form and matter is used to organize the modularity styles. But the same general mixing theme is the underlying topic of modularity itself – modularity architectures are about what is separated and what is mixed – and how it is mixed. At the broad framework level, the mixing of form and matter is always a *mixis*. The early computer science literature covers some of the possibilities. As already noted above, Dijkstra [21] focused on a complete separation. Parnas in *On the criteria to be used in decomposing systems into modules* [22] describes an aggregate of modules.

Modern computer application systems are sufficiently complex to require a network of processes for their development and operation. In this case, the modularity acquires a dynamic aspect that may extend over a connected series of mixing and separation processes. The final product may need to be a mixture of components, but the development may take a path through a variety of components mixed in different ways. The System Development Life Cycle (SDLC) in SSADM [23] and IDEF1X [24] are good early examples of this.

## 4      The context for the emergence of conceptual modeling

In this section we describe the information evolutionary context we adopted as our second framework. We found that an evolutionary context helps make sense of the emergence of conceptual modeling and its subsequent modularity adaptations. Evolution and modularity closely are linked. Wimsatt [25] has extended the work done by Simon [15] on complexity into evolutionary biology. One result is that he sees modularity as a "fundamental architectural feature" of evolution. Where modularity contributes to the manageability and evolvability of complex systems, as well as intersecting and complementing other fundamental architectural features such as robustness, reliability and entrenchment. This makes modularity a good lens with which to look at – and perhaps to guide – the evolution of conceptual modelling.

Though our specific interest is narrow – conceptual modelling – evolution has a deep relation to information that we tease out below. As we outline in the next section, the relation starts at the level of macro-evolution and extends across the whole of life. In the subsequent sections we look at the co-evolution of humans and information and the emergence of databases as the niche with the right conditions for the emergence of conceptual modelling.

### 4.1      The biological macro co-evolution of life and information

Evolution and information are intimately linked at the biological macro-evolutionary level. There is a pattern of information transitions that stretches across the whole of the macro-evolution of life. Maynard Smith and Szathmáry [26], [27], [28] frame it in these terms: "... that evolution depends on changes in the information that is passed between generations, and that there have been 'major transitions' in the way that information is stored and transmitted, starting with the origin of the first replicating molecules and ending with the origin of language." For them, these changes in information transmission, the passing of "information ... between generations", also known as inheritance, are the fabric of evolution. In [27], they provide a list of the eight major transitions. In each case, the transition not only transforms life but also transforms the way life evolves – and so, in a sense, evolution evolves through transformations in information transmission (inheritance). Though this is way outside the scope of modular architectural styles for computer applications, one can see a pattern of modularity styles here, where responsibilities are separated and coordinated.

### 4.2    The biological and cultural co-evolution of humans and information

Since Boyd [29] and others pioneering work in the 1980s, cultural evolution is now recognized as an established process within the broader framework of evolution. And in this wider framework, we can see that the same pattern of information transitions appears in human evolution – in both biological and cultural form. As Jablonka and Lamb [30], [31] note, the information transitions from speech, to writing, printing and, more recently digital computing are clearly major changes in the way that human information is stored and transmitted, where the earlier transitions more biological, the later one more cultural. These adaptations have accelerated and expanded human cultural evolution enabling more complex entities to emerge quicker, For more details on this see Ong [32] and Olson [33].

If we look at the pattern of evolutionary information transitions that led up to the latest human information transition, digital computing, there is an obvious feedback loop. One where each information transition constructs a 'niche' environment (a term from evolutionary biology [34]) that sets the stage for future evolutionary paths – including the next information transition. From this perspective, the evolution of the information ecosystem can be seen as an ongoing process of niche construction.

### 4.3    The emergence of databases – the niche for conceptual modelling

This broad evolutionary framework of information transitions evolving niche environments sets the broad background for the evolution of conceptual modelling, and more specifically the conceptual schema.

To set a more focused background we look at adaptations in the early stages of the latest human information transition, digital computing happening around the middle of the 20th century. This soon led to developments such as the programming of computer software and then the early databases. This opened a space into which early network and hierarchical data models could emerge. IBM developed the hierarchical data models, CODASYL (the Conference/Committee on Data Systems Languages) developed the network data model. This laid the foundations for the emergence of conceptual modelling.

We set the scene here, giving enough history of these early days to give a context for examining our topic. Most text books on databases (see, for example [3]) will give a little more detail on the history and references for further reading. By the late 1960s and early 1970s, hard disks became common, allowing any location on disk to be accessed in just tens of milliseconds. Data were thus freed from the tyranny of sequentiality. Though Codd [35] defined the relational model in 1970, it was not until the early 1980s that relational databases became competitive. And then soon replaced the earlier databases.

## 5     The emergence of conceptual modelling

The previous two sections described the two frameworks we are using to frame conceptual modelling's turn to schema. In the last section we described the emergence of databases which provided a fertile cultural (niche) environment for the emergence of conceptual modelling, as part of the ongoing evolution of the ecosystem. In this section, we look at the modular evolution of conceptual modelling. Intriguingly, there is an awareness of evolution running through the conceptual modelling literature – see [36], [37], [38]. In many cases, these look forward to future evolutionary trajectories.

### 5.1     Our narrow background

However, unlike most of these texts, we have a restricted focus on the emergence of conceptual modelling. Even here, there is rich history. Our purpose here is not historical exegesis, however much this is needed, but to capture the underlying patterns of modularity architectural style along a few broad dimensions, to create a broad picture of the landscape within which the conceptual modelling engineer has and can work. So, we focus on a couple of key features: the shifting use of the terms 'conceptual' and 'schema', the development of the conceptual model and its division into what came to be called the conceptual schema and the information base, particularly as it relates to the separation of the conceptual schema from the conceptual model.

There is a common view that, as Akoka et al [39] note "there is a reasonably common intuition about what conceptual modeling is". However, our closer look suggests that the notion has shifted over time across a range of senses. We look at three evolution threads that illustrate this. We start with a relational thread, then move onto an entity relationship thread and then finally look at an ANSI-SPARC inspired thread.

### 5.2     The early road to relational

One evolutionary thread was the development of the relational database. We sketch this history using three papers. We start with the seminal paper by Codd [35]. We then look at these two later CODASYL documents [40], [41] – a reasonably representative sample of the reports CODASYL produced at the time. What these three papers have in common is a focus on the database and no mention of the real world that is, presumably, being represented. There is mention in the two CODASYL documents of a conceptual framework, but no mention of a conceptual model.

In [35], Codd discusses concepts in a general sense, but nothing linked to a conceptual model. It also discusses data models – understandable given the paper's topic is the relational data model. There is an interest in data independence, but no mention of schema.

Both CODASYL documents describe the same Data Description Language (DDL) – at different stages of development. In neither document is there an explicit mention of a conceptual model, though there is talk of a 'conceptual framework' (Section *2.3 CONCEPTUAL FRAMEWORK* in the second document). Both documents have a diagram *1 – CONCEPTUAL DATA BASE MANAGEMENT SYSTEM* – though the details of the diagram are slightly different. The closest either of the two documents comes to mention that the data may represent something is where [40] says "Employ a data structure (s) that models the business or problem."

In both papers, it is made clear that the data description language is designed to be independent of any physical implementation, and the benefits this brings. Though the connection is not explicitly made, one could make a good case that the sense of conceptual here is 'independence from a particular physical implementation'. There is also a clear understanding of the distinction between data elements and the way they are organized into sets and relations – and that they are aggregated. The term 'schema' is defined as a set of DDL entries that provide a complete description of the database. The DDL entries define all the kinds of data in the database.

From our modularity architecture perspective, one can detect the CODASYL schema has a modular structure similar to that of a conceptual model. There is a material division of the conceptual framework from the physical storage devices, reflecting the aim of providing data independence to the CODASYL schema. There is a formal connection (aggregation) in the mappings between the physical devices' data to the schema. The various elements of the CODASYL schema (and the relational model) are connected. But there is no further modular division of the conceptual framework into a conceptual schema and information base.

### 5.3   The road to entity-relationship modelling

Alongside the work that focused on the databases themselves, there is another evolutionary thread that considered what the data in the database represented which led to entity-relationship modelling. We sketch this history using two key papers by Mealy [42] and Chen [43]. The papers take broadly similar positions as the brief summaries below show. What is interesting for us is that they both introduce a distinction between a 'data' model that represents the real world and one that represents the 'machine data systems'.

In [42], Mealy uses the term 'concept' in a general sense, but the only mention of 'conceptual' is a reference to the 'conceptualist' philosophical position as part of a more general discussion of ontology. He writes about three systems "The first system is … some part of the real world, the second is our theory of the first, and the third is a machine representation of that theory." He continues "A representation is, itself, now defined as a map establishing a correspondence between two systems."

He writes "We have called data fragments of a theory of the real world." He makes a clear distinction between types of things (in the real world) and how this is reflected in the model: "The model is a system of sets of entities, values, data maps, and procedure maps. The entities correspond to the objects in the real world about which data are recorded or computed." The types of things (a material division) are (formally) aggregated (using relations) but there is no additional material division of the model into anything resembling the conceptual schema and information base. This is no mention of schema, but a mention of data descriptions as the language for this model and noting these should be machine independent.

In [43], Chen states in first few sentences of the abstract that his "model incorporates some of the important semantic information about the real world." He writes of "2.2.4 Conceptual Information Structure" – but not conceptual models. There is a brief mention of schema – as data definition – but this encompasses the whole model. Within the model there are (material) distinctions between entities and sets (formally) aggregated by relations – but there is no additional (material) division into conceptual schema and information base. He makes a (material) distinction between representations of the world (via concepts) and representations of information – levels 1 and 2.

From our modularity architecture perspective, the first key modular structure is the notion of logical views. This makes a material distinction broken into four levels, of which the first two are the most interesting. The first level is *Information Concerning Entities and Relationships*; though Chen does not specifically do this, we can consider this a conceptual model. The second is *Information Structure* – which considers 'representations of conceptual objects'. This has a layered formal modularity, where the level 2 information structure view is an integration (so a *mixis*) of the information structure with the information view.

In Section *3.2 An Example of a Database Design and Description*, Chen introduces two more important (for us) modularity architecture styles. He notes that "[t]here are four steps in designing a database using the entity-relationship model. In the first step he restricts the modelling to "the entity sets and the relationship sets of interest", in the third step he "define[s] the value sets and attributes" and in the fourth step he "organize[s] data into entity/relationship relations and decide primary keys." These imply some related architectural styles.

Firstly, it introduces a three-way material division between the entity-relation sets, the value sets and attributes and the data. Secondly, it introduces the formal modular style that produces a separation followed by a dependent separation – which when 'mixed' with the material division produces a concrete modularity style. In this concrete style, the instance data and value set-attribute levels are separated from the set level and set aside for later integration. Then firstly the value set-attribute level and then subsequently the instance data levels are integrated in turn in situ.

The technology of the 1970s (and much of the 1980s and 1990s) could not scale easily to handle the modeling of the volume of large systems' instance level data – even though systems then current were not large by modern standards. So, the focus on schema was making a virtue of technical necessity.

The technology of the 1970s even had issues with modelling the schema level data, so Chen's sub-divisions using the two-level material division reduced the technology burden further. It enabled a greatly simplified first task that could focus on the broad structure undistracted not just by the details of instance data, but also of value sets and attributes. Then the third task can focus on value sets and attributes, undistracted by the details of instance data. However, a point we shall return to later, this presupposes the practicality of a rational approach to design. One where we can work out the architecture in our head from first principles sufficiently well so that we do not need to empirically test our design against the data.

### 5.4   An ANSI/SPARC inspired ISO 'standard'

The final evolutionary strand we look at follows the ANSI-X3-SPARC Study Group. We look at the history through the lens of two papers; an ANSI-X3-SPARC report [44] that culminates in an ISO Technical report – *Concepts and terminology for the conceptual schema and the information base* [1] – which ISO reviewed in 2022 and confirmed as current.

The central architectural structure in the ANSI report [44] is an architecture composed of three levels (external, conceptual and internal) which map onto each other. The conceptual model is "a collection of objects that represent the entities in an enterprise." This is composed of a conceptual schema and conceptual data – where the schema is bound to the data.

The report suggests that the external and internal views were (at the time of the report – 1975) a common two-level structure for databases. It makes the case that the third conceptual level is usually informally invoked to settle disputes between the original two levels – and that "it must be made explicit and, in fact, made known to the data base management system. The proposed *mechanism for doing* this is the conceptual schema." Where conceptual data is bound to the conceptual schema.

The ISO Technical Report [1] builds upon the ANSI-SPARC work. The report starts (in section 1.1. THE ANSI/SPARC FRAMEWORK) by referring back to the two ANSI/SPARC reports [44], [45] – as well as referring to subsequent work.

It adopts the name 'universe of discourse' for the portion of the real world being represented by the conceptual model – so clearly recognizing a material distinction of conceptual models.

Stating that "within the context of an ANSI/SPARC framework: We consider both the conceptual schema and the information base to be at the conceptual level, providing a conceptual view of the information about the universe of discourse." Recall the prior ANSI-SPARC report reports [44] used the term conceptual data rather than information base.

It notes that sometimes in the literature "no clear distinction is made between the things and the description of the things, nor between the information meaning and the data representation." And that this sometimes rests on a fundamental disagreement whether the structure of the conceptual is dependent upon the entities in the universe of discourse or upon the generic data structures being used. This prefigures current disagreements about the use of top ontologies and generic data models. It makes clear it takes the view that the conceptual model represents the real world. This, it can be argued, gives it clear independence from any specific data implementation. For simplicity, we follow the same course, though we return to this point later.

It considers both the syntactic and dynamic aspects of the conceptual model. It notes that "[m]uch of the past work on concepts for the conceptual schema has been concentrated on the static aspects, … the set of concepts for the conceptual schema should also cover the dynamic aspects. A point we shall return to later is that it, in terms of our modularity framework, recognizes a material division between static and dynamic – one that should be included in the conceptual model.

It makes the scope of the conceptual schema clear, stating that it "describe[s] classes (types, variables) in the universe of discourse rather than individuals (instances)". It also states that "[t]he main reasons for this distinction are practical arguments with respect to the design, implementation and maintenance of the information system".

From our modularity perspective, we now have a position where the material distinction for the conceptual model is that it represents the universe of discourse. We have a further distinction that divides the conceptual model into the conceptual schema (classes) and information base (instances). We have a two-stage concrete architectural style that at the start of development separates the information base – to allow the building of a separated conceptual schema. Followed by a later merging in of the information base into the pre-built conceptual schema.

With hindsight, one can see the choice of this style was guided by the state of the information ecosystem. Firstly, much development was greenfield. This meant that the information base was not easily available in a machine-readable form – one suitable for automatic consumption. Secondly, the tools that supported the development of the conceptual schema could not easily scale to consume a much larger information base. The choice of this style makes sense in the context.

## 6    The emerging formal and material hierarchy

The content of the last section provides resources for us to flesh out some of the modularity framework. We have a range of concrete architectural patterns from which we can extract the obvious formal and material elements. In the last section, we saw a two-stage concrete style which starts with the building of the conceptual schema and then the information base is merged in. Formally, this is a separation of the conceptual model into two, where one part is built and then the other is merged in. Materially, it involves two distinction styles. Firstly, a distinction of the conceptual model and then a division of this into a conceptual schema and an information base.

In earlier sections, we saw holistic versions of the conceptual model – where the conceptual schema and information base are aggregated. This involves the same material distinctions but a different formal modularity style. We also saw concrete styles involving conceptual and physical models. Given the way they were described, a reasonable way of interpreting the physical models is as an integrated mixture of the conceptual model and its physical implementation. This illustrates the formal integration style and the conceptual-physical material distinction style.

### 6.1    The emerging hierarchy

The information ecosystem has evolved over the decades since the conceptual model and its division into the conceptual schema and information base emerged. We need to refine the hierarchy to accommodate that evolution – mostly evolving the material distinctions – to deploy in the current environment.

We firstly need to refine and generalize the conceptual-physical material distinction style. We make the distinction: conceptual versus implementation. Where conceptual focuses on what the systems' relation with the universe of discourse is, and implementation focuses on how the conceptual system is implemented. The distinction is not totally straightforward, as any conceptual system will be implemented in some way. The notion of implementation in play here is where the choice for the implementation of the target system is made. A conceptual 'implementation' avoids commitment to the target system having a particular implementation. Within the conceptual level, there is also the material distinction, already discussed, between (conceptual) schema and (conceptual) base.

Within the conceptual level, there is a material distinction between ontology and agentology. Ontology here has the traditional sense of Nagel's [46] 'a view from nowhere' and agentology is those features that are indexed to the system. Alternatively, it can be described as a distinction between a *de re* (ontology) and *de se* (agentology) view. The papers [47] and [48] give a detailed description of this distinction. They also describe two possible concrete modularity architecture styles – the traditional *mixis* architecture in most modern applications, where ontology and agentology are closely entangled and a layered style, where agentology is a distinct layer on top of ontology.

The ISO Technical Report – *Concepts and terminology for the conceptual schema and the information base* [1] – introduced a clear requirement for handling both static and dynamic aspects. In the conceptual sphere, the dynamic aspect is the agent's changing relationships with the world. The papers mentioned above, [47] and [48] have good examples of this.

We can discern two ways the agent's relationship with the world can change. These provide us with the next material distinction: doxastology (epistemology) and deontology. Where doxastology (epistemology) handles the changes in belief (knowledge) that the representation records. And deonotology, the changes that the agent makes to the world. Austin [49] and Searle [50] notion of the illocutionary force of speech acts provide a neat illustration of this. They also introduce the notion of direction of fit – world to word and word to world – which is one way of distinguishing doxastology and deontology. This distinction is not currently well-recognized, so the possible modularity styles are not well studied.

Within the implementation level there are various overlapping material distinctions. There is a distinction between structure and behavior (closely related to that between static and dynamic). Note that the same conceptual item – for example, the balance on an account – can be implemented as stored structure and or as a calculated process (behaviour). Stored procedures are a good general example of this kind of duality. This is what makes the distinction an implementation rather than a conceptual decision. Note also, that the implementation of the conceptual system typically includes aggregated modules of both structure and behavior.

There is a distinction between the different broad types of physical substrate the system is implemented on. Simplistically, whether it is implemented using humans or machines. Often it is implemented on a combination of both – in our modularity architectural terms, in two sub-modules, one human one machine. When fully automated, it is implemented upon just machines – so architecturally one machine module.

There can also be, at the implementation level, a distinction between the physical schema and the physical base. This does not have to track the corresponding conceptual distinction, mentioned earlier, though for simplicity people usually assume it does.

## 7      An example of a concrete inclusive schema-and-base modularity architectural style

Earlier in the paper we noted that in the 1970s and 80s when conceptual modelling emerged, there were (niche) environmental constraints the encourage schema-based conceptual modelling. The lack of scalable technology meant that the size of the conceptual model had to be kept small. There were also many more greenfield projects where suitable information base data was not easily available. So, there was an understandable focus on conceptual schema – often with a restricted scope – and the information base was separated and its amalgamation put on hold until much later in the development process. The information ecosystem has evolved significantly since then and these constraints have unraveled. However, as noted earlier, the habits associated with these constraints are so entrenched that they are still relatively unquestioningly observed.

We expose these entrenched habits with an example where the modularity architectural landscape suggests an inclusive schema-and-base modularity architectural style makes sense. Our example is a common task in our contemporary information ecosystem, package application migration (which can be considered a specific case of Architecture-Driven Modernization (ADM)). In doing so, we open the general possibility that in the overall modularity architecture landscape there may be many other new ways of conceptual modelling waiting to be surfaced.

As packaged application migration can vary considerably, we need to be more specific. Migrations can be between versions of the same base package – like, for example, the current SAP HANA upgrade. Or they can be between different packages. They can provide opportunities for implementing substantial changes or not. The migration between different packages with no changes illustrates the situation better, so we choose this for our example.

### 7.1      Package application migration example

We start by considering the migration as a black box with a source (input) and target (output) package. The migration is essentially a movement of the base from the source package to the target package. This becomes even clearer if one considers a series of package application migration. What persists through these is the base – at each migration the source physical schema is dropped – replaced with the target schema. The obvious modularity pattern for these packages in the architecture uses a physical schema and base material distinction – formally mixed.

A biological analogy helps to reinforce this point. In the nineteenth century, Weismann [51] noticed that the mechanisms of genetic transmission typically only transmitted some types of information. He made a basic (since refined) material division of cells into the germline and the somatic line. The germline is those cells that are involved in reproduction and the transmission of genetic information from one generation to the next, they are heritable. The somatic line is the rest of the cells, the non-reproductive cells. They are not involved in the same way in transmission, so they are non-hereditary. The germline cells have been called 'immortal' in the sense that they (or their genes) continue to exist indefinitely through reproduction – creating a lineage. Whereas the individuals and their somatic line cells die, they are mortal. Dawkins' replicators and vehicles [52] can be seen as a generalisation of this, with replicators in place of germlines and the vehicles in place of soma.

The physical bases (the base-data-line) are analogous to the germline (or replicators); they are what persists persist between generations. The schemas (the schema-line) are the somatic lines (or vehicles); they do not. The analogy is not exact. In the package application migration architecture, the target system physical schema (analogous to the new somatic line) is pre-constructed and then populated. In the genetic, Weismann case, the new somatic line grows based upon the germline. However, this does not detract from the point that analogy highlights, that the physical base is central to the migration.

There is one further material distinction to clarify. In the current ecosystem, most source and target systems are largely automated – so implemented in the 'machine' space – and the volume of physical base data is significant. This suggests that the core of any migration needs to be automated – an architectural style in the machine side of the human-machine material divide. Of course, the processes for this automated migration may need to be manually developed before the migration process can be run.

In practice, the source physical schema in not discarded. The source physical base often contains substantial amounts of data, far too much to be migrated individually, whereas the physical schema has orders of magnitude less data. So, the natural next step is to use the source and target physical schemas as the scaffolding for algorithms to automatically map the source physical base data to the target.

This mapping naturally raises the architectural question of what material difference the mapping is over. A natural answer, given the conceptual modelling literature (particularly the Mealy [42] and Chen [43] branch) we have reviewed, and our modularity framework, would be that the source and target systems are both implementations of a common conceptual model, one which shares a common universe of discourse. And each of the physical systems is a mixture of the conceptual model with the physical design of the system. From this perspective, the mapping essentially removes the source physical design, preserving the core conceptual system and adds in the target physical design.

This insight enables us to articulate a modularity design choice. If the physical systems share a common conceptual model – then the mapping could be easier and of a higher quality, if there was a two-stage mapping – firstly into the conceptual model and then to the new physical implementation. Where the first mapping is an unmixing and the second a mixing.

Often, in practice, the next architectural decision is schema-centric – to adopt a modularity architecture that separates the schema from the base and creates two scaffolding mappings between the three schemas. And then afterwards use these mappings to map the base twice. This is what the 'shifting left' [53] literature calls a shift right choice – and it makes clear this typically has the consequence that conceptual mistakes in the mapping will only be found as they are volume tested, late in project, with all the attendant increased costs.

The modularity architecture landscape is clear that there is another shift left choice we could make, one which avoids these shift right costs. That is a style with a single module for the physical system – an inclusive schema-and-base module – and a process that maps it as a whole.

## 8    An implementation of the expansionary example

In the previous section, an example of type of perspectives that the modularity architecture can open up was explored – one where the information base is not separated from the conceptual schema – one that treats the model (and so, conceptual model) as an inclusive whole. We have now been working for decades with this kind of modularity architectural style, implementing it in a data pipeline based upon a framework now called bCLEARer. Over time we have learnt some practical lessons, which we share here. We start by outlining the context, starting with the historic tool context and then moving to the bCLEARer context.

### 8.1    Historical tools context

The requirements of schema-and-base conceptual modelling, with its need to incrementally transform a typically sizeable information base, naturally leads to data pipeline tooling. To appreciate the radical differences this approach brings, it is useful to contrast it with the historic tools that emerged after the schema turn.

In the 1970s, the early CASE tools were diagramming aids for capturing specifications in structured graphical form. In the 1980s these developed into integrated repositories offered commercially, such as IEW (Information Engineering Workbench) and Excelerator (from Index Technology). These offered manual graphically based editing that focused on the schema – clearly schema biased. UML emerged as a kind of common standard in the mid-1990s – but UML tools still offered GUI-based editing of a common repository storing the schema as data.

In the late 1990s and early 2000s, Model-Driven Engineering (MDE) and textual Domain-Specific Languages (DSLs) emerged. These has less reliance of GUI editing – especially text-based DSLs – and included transformations as part of the process. The usefulness of visualization was recognized in graphical projections of the models [54]. Nevertheless, these tools were still schema based (schema-biased). Furthermore, the practices typically started with manual textual or graphical conceptual modelling, with the subsequent (automated) transformations being life cycle focused. This contrasts with the data pipeline ETL (Extract, Transform Load) frameworks, such as Apache NiFi or Talend Open Studio, where the modelling process itself is automated.

### 8.2  bCLEARer context

For bCLEARer context, we start with a brief description of the framework and then move onto the lessons learnt. bCLEARer was initially developed and deployed in the late 1980s, this early work is described in *Business Objects* [55]. It was originally developed to address a particular need for a solid legacy reengineering process. This naturally led to the development of a methodology for re-engineering existing information systems, currently named bCLEARer – where the capital letters are an acronym for Collect, Load, Evolve, Assimilate and Reuse. This was co-developed with a closely intertwined top-level ontology (the BORO Foundational Ontology).

At the time these were initially developed (the late 80s and early 90s) the technology support was primitive, so while the core process was systematized it was not automated. Over the last few decades, as appropriate technology has emerged, the core process has been fully automated in a data pipeline [4]. Later versions of this are described in several places, including [56], There are also open-source examples on GitHub: github.com/boro-alpha/bclearer_boson_1_1, github.com/boro-alpha/bclearer_boson_1_2 and github.com/boro-alpha/uniclass_to_nf_ea_com. In its current state, bCLEARer provides both the framework and the data platform for implementations of modularity architectures that undertake conceptual modelling (schema and base) automatically in a data pipeline.

bCLEARer (and its associated top-level ontology) have, over the years, been configured to exploit a variety of situations ranging from its original legacy system migration to application migration to developing requirements and quality controlling existing systems. A common feature of all these projects was the initial collection of one or more datasets and their regimented evolution to a more digitally aligned state. Often this was an ontology, a kind of conceptual model.

### 8.3  Lessons learnt

Working on these projects, we have found that adopting this modularity architectural style leads to changes in practice – and our view of what the modelling we are doing correspondingly shifts.

Conceptual modelling is currently typically a human working at a screen updating the conceptual schema model repository – whether graphical or textual. From the modularity architectural style perspective, this is a hybrid aggregation of human-machine modules. When one adds the typically sizeable information base to the model, this architecture style is much less feasible due to scaling issues. Modelling changes to the schema now need to be rolled out across the information base and this adds a significant cost. It is only easily achieved if undertaken by a machine, in other words automated. In terms of the human-machine material distinction, this changes the core modularity architectural style to machine-only. Some preliminary coding may be done externally by humans, but the core modelling is done by a machine in the data pipeline.

This then created pressure to adopt engineering practices suited to data pipelines common in data engineering. Good observability and inspectability are needed to follow this conceptual modelling [57]. The automation then drives down the costs of transformation.

Current conceptual modelling practice can be characterized as rational – where the decisions are rationally assessed and the appropriate conceptual schema is built. Further rational analysis may give rise to changes in the conceptual schema. In this scenario, our judgment plays a deciding role. At the end of the development process, we introduce the information base and so can test the schema. When we adopt an inclusive schema-and-base style (and automation has kept the costs under control), we have a new empirical possibility from the start, we can now test the changes in the conceptual schema against the information base.

Kirsh and Maglio's *On Distinguishing Epistemic from Pragmatic Action* [58] found that certain cognitive problems are more quickly, easily, and reliably solved by performing actions in the world than by performing computational actions in the head alone. That sometimes – what Kirsh and Maglio call – epistemic actions (actions performed to uncover information that is hidden or hard to compute mentally) are more effective than pragmatic actions (performed to bring one physically closer to a goal). We found something similar with our pipeline modelling. Provided we could keep the cost of transformation low, it was often more effective to test modelling decisions against the information base data than attempt to rationally analyze it. A key factor in managing the cost of transformation was the development of appropriate tools. The MDA literature (e.g. [59], [60]) reports a similar story about automation and tools. In the case of data pipeline conceptual modelling, the data engineering community has already developed a useful toolset that provides a good foundation.

Even as early as Mealy [42] in the conceptual modelling literature a story developed of 'three worlds' – three modules – the real world, the description of the world and the description of the system describing the world. One sees the same 'story' in the database development migration from the Conceptual Model (Mealy's world 2) to the Logical Data Model (Mealy's world 3).

Similarly, a distinction was made between models that simulated what they were modelling – exemplifying some of the properties of what was being modeled and models that instead described it. One can see this as belonging to a wider distinction between showing and telling. Macbeth [61] both shows and describes how this same distinction appears in mathematical proofs. She gives, as a simple example, the comparison of seeing the arithmetic of Arabic numerals and with the description needed for Roman numerals. Under the standard story, in Mealy's three worlds, the one-to-one representation relation between worlds plays a 'telling' function – albeit one based upon, as Mealy points out, on an isomorphism between the represented and the representation.

However, close acquaintance with the full conceptual model in a data pipeline suggests another plausible story. In the data pipeline, the transformation from the (whole) conceptual model to the logical model does not feel like a shift up a level of representation – it rather feels more like an expansion. This suggests a different story, one where there are only two worlds (two modules), the real world and then a second world that has both a showing and a telling function – whether it is the conceptual model or the logical model. It tells about the real world and shows the system that is doing the telling. The logical model is just adding more showing. This suggests a re-interpretation of both the database development and MDA stories – where the models are not descriptions of systems, but early prototypes of the system that evolve into the final system.

## 9      Future work

There are two strands of possible future work; one which focuses on the short-term goal of exploiting the possibility of holistic schema-and-base conceptual modelling, the other a longer-term goal of finding other architectural possibilities for deploying conceptual modelling.

For the first short-term goal, there is work to be done to refine the current approach, especially the automation tools. Despite their schema-centricity, useful role models here are MDA [59], [60] and CMP [62] – especially their adoption of machine/automation. We need to learn what lessons we can from them to add to the tools data engineering has brought.

For the longer-term goal, we need to develop the modularity framework, to help us identify the opportunities better. This will involve both refining the hierarchies and developing more example architectures to illustrate their deployment. Some useful work could probably be done on the taxonomy of modular mixing.

## 10      Summary

We have developed a framework for examining the modularity architectural styles in the evolution of conceptual modelling. This has enabled us to look back at conceptual modelling's early development and see the emergence of the schema turn – where the conceptual modelling completely separates the information base from the conceptual schema and sets it aside. We identified the niche pressures – greenfield projects and under-powered, poorly scaling technology – that made the turn a sensible adaptation.

We then showed that the turn is not fundamental. We revealed it as possibly a temporary evolutionary detour with modern technology enabling the adoption of an inclusive schema-and-base conceptual modelling approach. We did this in two stages. We firstly showed how an examination of the molarity landscape for a common task, package application migration reveals a possible style involving an inclusive schema-and-base style. We then described our experience working with this, using the bCLEARer method noting some lessons learnt. These included increases in the level of automation and the ability to design empirically (epistemic actions). We hope this illustrates that the space of possible conceptual modelling practices could be wider than currently assumed, that there could be a range of new ways of using conceptual modelling to address current information system requirements.

**Acknowledgments.** We would like to acknowledge the help of Mesbah Khan, Jonathan Eyre and Philip D'Rozario in the development of this paper.

**Disclosure of Interests.** The authors have no competing interests to declare that are relevant to the content of this article.